\documentclass[aps,pra,reprint,superscriptaddress,longbibliography,floatfix]{revtex4-2}
\usepackage[T1]{fontenc}
\usepackage[utf8]{inputenc}
\usepackage{amsmath,amssymb,bm,graphicx,microtype,booktabs}
\usepackage{xcolor}
\usepackage[colorlinks=true,allcolors=blue!45!black,bookmarksnumbered=true]{hyperref}
\hypersetup{pdftitle={One Bit, Many Records: Information Replication in Quantum Many-Body Dynamics},pdfauthor={}}
\graphicspath{{figures/}}
\newcommand{\Tr}{\operatorname{Tr}}
\newcommand{\ket}[1]{\lvert #1\rangle}
\newcommand{\bra}[1]{\langle #1\rvert}
\newcommand{\eps}{\epsilon}
\begin{document}
\title{One Bit, Many Records: Information Replication in Quantum Many-Body Dynamics}
\author{Dan-Bo Zhang}
 \altaffiliation[Contact author: ]{dbzhang@m.scnu.edu.cn}
 \affiliation{Key Laboratory of Atomic and Subatomic Structure and Quantum Control (Ministry of Education), Guangdong Basic Research Center of Excellence for Structure and Fundamental Interactions of Matter, and School of Physics, South China Normal University, Guangzhou 510006, China}
 \affiliation{Guangdong Provincial Key Laboratory of Quantum Engineering and Quantum Materials, Guangdong-Hong Kong Joint Laboratory of Quantum Matter, and Frontier Research Institute for Physics, South China Normal University, Guangzhou 510006, China}
 \author{Yan-Ming Chen}
 \affiliation{Key Laboratory of Atomic and Subatomic Structure and Quantum Control (Ministry of Education), Guangdong Basic Research Center of Excellence for Structure and Fundamental Interactions of Matter, and School of Physics, South China Normal University, Guangzhou 510006, China}
\begin{abstract}
Information replication occurs when a single locally encoded classical bit becomes simultaneously readable across extensively many disjoint regions of a quantum many-body system. Here we identify dynamical selection as a mechanism for generating these records, whereby an unperturbed nonthermal reference remains locally recognizable while its locally perturbed counterpart spreads and relaxes under the same Hamiltonian. We establish an entropy-capacity criterion showing that a uniform area-law reference paired with a volume-law perturbed branch guarantees extensive records at a fixed, nonzero information threshold. An embedded-dimer chain demonstrates both record proliferation and increasing readout reliability with fragment size, while stabilized PXP dynamics extends the mechanism to a coherently recurrent reference. These results characterize information spreading through the spatial multiplicity of accessible records and provide experimental signatures measurable via fixed local readouts.
\end{abstract}
\maketitle
\section{Introduction}
\label{sec:introduction}
Quantum information dynamics characterizes how interacting many-body systems respond to local perturbations, from quasiparticle transport~\cite{Calabrese2005} to operator spreading and scrambling~\cite{Nahum2018,Keyserlingk2018,Khemani2018}. Its characterization draws on entanglement growth~\cite{Kim2013,Nahum2017}, out-of-time-ordered correlators~\cite{Maldacena2016,Garttner2017,Xu2019}, operator-size distributions~\cite{Roberts2018,Qi2019,Hu2023,Blocher2024}, and scramblons~\cite{LiuZhang2024Scramblon,Li2026Scramblon}. Complementary probes include Holevo-information dynamics~\cite{Yuan2022,Zhuang2022,Sun2025}, subsystem information capacity~\cite{Chen2025SIC}, and randomized measurements~\cite{Vermersch2019,Joshi2020,Chen2026}.

While familiar diagnostics capture the propagation and redistribution of quantum information, they do not directly establish whether a single local event leaves readable evidence in many places at once. A quasiparticle can transport information without leaving copies behind, and scrambling can hide information in nonlocal correlations inaccessible to small fragments~\cite{Hayden2007,Hosur2016}. This distinction motivates a complementary question of whether a single locally encoded classical bit can become simultaneously accessible to extensively many spatially separated observers. We term this phenomenon \emph{information replication}. It measures the spatial multiplicity of accessible records rather than the causal reach of a disturbance or the information recoverable from a single optimized fragment. Individual records may remain incomplete and mutually correlated, while the initial input still encodes only one classical bit.

Achieving replication is nontrivial in isolated dynamics. A local operation alters global conserved quantities by only $O(1)$, so two nearby preparations share identical limiting energy and charge densities. If both states relax to the same local thermal ensemble, their initial distinction becomes inaccessible to local observers despite being preserved globally by unitarity~\cite{Deutsch1991,Srednicki1994,Rigol2008}. Retaining a recognizable reference while allowing the disturbance to reach distant regions requires dynamics sensitive to the initial preparation. Embedded nonthermal eigenstates and quantum many-body scars offer a natural route~\cite{Shiraishi2017,Turner2018,TurnerPRB2018,Moudgalya2018,Choi2019}, where a protected state remains nonthermal while a local perturbation activates spreading and relaxation. Macroscopic responses to local perturbations in kinetically constrained systems provide related examples of such state-dependent evolution~\cite{Bidzhiev2022,Fagotti2024}.

\begin{figure}[!t]
\centering
\includegraphics[width=\columnwidth]{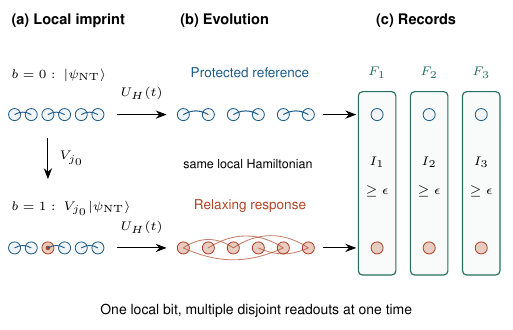}
\caption{\label{fig:concept}One local imprint yielding many simultaneous readouts. A classical bit selects either a protected nonthermal reference or its local perturbation. Both branches evolve under the same local Hamiltonian. The reference remains recognizable while the perturbed branch spreads and relaxes, producing distinct local outcome distributions on disjoint fragments $F_a$. Each informative fragment records the same initial event. The sites and correlations are schematic, and no specific spin order is assumed.}
\end{figure}

We identify this dynamical selection as a general mechanism for information replication [Fig.~\ref{fig:concept}]. Evolving under the same Hamiltonian, the reference preserves a recognizable local pattern whereas the perturbed branch explores configurations outside that pattern. Observers can then detect the initial operation through local departures from the reference, and relaxation within the perturbed branch sustains this contrast long after the microscopic details of the seed have dispersed. Crucially, spatial propagation and record formation impose distinct requirements, as the disturbance must both reach distant regions and leave locally readable evidence at a common observation time.

Our theoretical framework connects this local distinguishability directly to the entanglement structure of the two conditional branches. A reference obeying a uniform area law has a bounded local entropy capacity, whereas a volume-law perturbed branch forces extensively many local regions to explore states outside that reference support. This mismatch guarantees an extensive number of partial records above a nonvanishing information threshold. We demonstrate this mechanism using an exactly protected singlet reference in an embedded-dimer chain and extend it to a coherently recurrent scar reference in the PXP model. The main text develops the physical framework and core demonstrations, while Appendices~\ref{app:information}--\ref{app:pxp} provide formal proofs, model specifications, and numerical verifications.

\section{Local records and entanglement contrast}
\label{sec:theory}
\subsection{Information available to disjoint observers}
Consider a pure reference state $\ket{\Phi}$ on a spin chain. A single classical bit $B$ is encoded locally such that $B=0$ leaves $\ket{\Phi}$ unchanged and $B=1$ applies a bounded-support unitary $V_{j_0}$, preparing $\ket{\psi_0(0)}=\ket{\Phi}$ and $\ket{\psi_1(0)}=V_{j_0}\ket{\Phi}$. In our concrete models, the two initial states are orthogonal and evolve under the same local Hamiltonian according to $\ket{\psi_b(t)}=e^{-iH_Lt}\ket{\psi_b(0)}$. An observer monitoring a bounded fragment $F$ accesses the reduced state $\rho_{b,F}(t)$. For a specified local measurement $\mathcal M_F=\{M_y\}$, we define the conditional outcome probabilities $p_b(y)=\Tr[M_y\rho_{b,F}]$ and the marginal distribution $\bar p=(p_0+p_1)/2$. The accessible mutual information is
\begin{equation}
 I_{\mathcal M}(F,t)=H(\bar p)-\tfrac12H(p_0)-\tfrac12H(p_1),
 \label{eq:mutual}
\end{equation}
where $H(p)$ denotes the Shannon entropy in bits. While bounded from above by the Holevo quantity according to $I_{\mathcal M}\le\chi_F\le1$, $I_{\mathcal M}$ directly quantifies what a chosen experimental readout recovers~\cite{Holevo1973,Zwolak2013} (see Appendix~\ref{app:information} for the measurement-independent comparison).

Operationally, the conditional distributions $p_b(y)$ are calibrated by repeatedly preparing each branch. An observer presented with a single run of an unknown preparation then extracts statistical evidence about the input bit from the local measurement outcome. In practice, the reference pattern dictates a natural readout, namely singlet-projector tests for embedded dimers and laboratory-$Z$ measurements for the PXP chain. Within each model, all observers share the same fixed local measurement rule independently of the unknown bit label, eliminating the need for position-dependent basis optimization.

The information threshold $\eps$ specifies the minimum evidence obtained by each observer without requiring error-free identification in a single shot. Distinct fragments may carry noisy, mutually correlated records of the same \emph{classical} input bit. Because this process replicates a classical label rather than an arbitrary unknown quantum state, it fully respects the no-cloning and no-broadcasting theorems~\cite{Wootters1982,Barnum1996,Piani2008}. Counting readable fragments cleanly separates the spatial multiplicity of records from the single-shot reliability of any individual fragment.

For a fixed family of bounded fragments and local readouts, let $R_\eps(L,t)$ denote the maximum number of mutually disjoint fragments satisfying $I_{\mathcal M}(F,t)\ge\eps$. Extensive information replication is defined by
\begin{equation}
 \liminf_{L\to\infty}\frac{R_\eps(L,t_L)}L>0
 \quad\text{at fixed }\eps>0,
 \label{eq:extensive}
\end{equation}
where all observers measure at the same time $t_L$ and the maximum fragment size remains independent of $L$. Enlarging the fragment size can enhance the reliability of each individual record at the expense of accommodating fewer disjoint observers; when testing extensivity at a target accuracy, the fragment size is held fixed as $L$ increases.

We focus on linear observation clocks $Jt_L=cL+O(1)$, where $J$ is the microscopic interaction scale and $\hbar=1$. For Hamiltonians with bounded interaction range and strength, Lieb--Robinson locality requires a propagation time $t=\Omega(L/J)$ before a local imprint can become readable above a fixed threshold across a finite fraction of a one-dimensional chain~\cite{Lieb1972,Bravyi2006}. Linear time therefore serves as a natural causal target. Note that this propagation bound does not constitute a general upper bound on relaxation times, since actual equilibration depends sensitively on the initial state, observable, and spectral gap structure, and may require longer timescales~\cite{Short2012,GarciaPintos2017,Heveling2020}. The bounded $O(1)$ clock offset allows one to align the readout with a reference revival peak. Our numerical calculations below test this linear timing across accessible finite system sizes.

\subsection{Main result and its physical explanation}
\noindent\textbf{Main result (informal).}
Suppose a single classical bit is encoded via a bounded-support operation on a pure reference state, and both preparations evolve under the same local Hamiltonian. If, at a common observation time, the reference branch obeys a uniform interval area law while the perturbed branch exhibits a nonzero volume-law entanglement-entropy density across a macroscopic interval, then a positive density of disjoint, fixed-size regions can each reveal at least a fixed, nonzero amount of information about the initial bit. The fragment size, information threshold, and record-density bound are all independent of system size (see Appendix~\ref{app:entropy} for the formal statement and proof).

Physically, this result stems from a local entropy-capacity mismatch. A short-range entangled reference state occupies only a low-dimensional subspace within each local block. Because the entropy of a composite interval is bounded from above by the sum of its constituent block entropies, a volume-law perturbed branch cannot remain confined within those restricted reference subspaces on almost every block. Consequently, a finite fraction of blocks must acquire substantial probability weight outside the reference support, allowing local projective tests to detect the initial event across multiple spatial regions.

A pure product reference illustrates this mechanism transparently. Suppose the reference factorizes into pure elementary cells of Hilbert-space dimension $D$ with local states $\ket{\phi_a}$, where each cell may be internally entangled like a two-spin singlet. Performing the binary projective test $\{\ket{\phi_a}\bra{\phi_a},1-\ket{\phi_a}\bra{\phi_a}\}$ accepts the reference branch ($b=0$) with certainty and rejects the perturbed branch ($b=1$) with departure probability $q_a=1-\bra{\phi_a}\rho_{1,a}\ket{\phi_a}$. The resulting mutual information is
\begin{equation}
 I_a=f(q_a),\qquad f(q)=h_2(q/2)-\tfrac12h_2(q),
 \label{eq:f}
\end{equation}
where $h_2(x)$ is the binary Shannon entropy. Because $f(q)$ increases monotonically with $q$, any finite departure probability immediately yields a finite local record.

A cell that rarely departs from its reference state has a strictly limited entropy capacity, as it resides predominantly in a single pure state with only $D-1$ orthogonal states available to absorb the remaining weight. Specifically, pinching $\rho_{1,a}$ in the reference test basis shows that (Appendix~\ref{app:entropy})
\begin{equation}
 S(\rho_{1,a})\le h_2(q_a)+q_a\log_2(D-1)\equiv g_D(q_a),
 \label{eq:capacity}
\end{equation}
where the first term accounts for the binary decision of leaving the reference state and the second bounds the entropy within its $(D-1)$-dimensional orthogonal complement.

Because any single cell can contribute at most $\log_2D$ bits of entropy, a sparse set of strongly disturbed cells cannot account for an extensive interval entropy. Crucially, our hypothesis bounds the global entanglement entropy $S(\rho_{1,A})$ of the entire interval $A$ from below, rather than merely the sum of individual cell entropies, which can be much larger in the presence of inter-cell correlations. Entropy subadditivity therefore certifies an extensive number of readable cells without requiring knowledge of their spatial locations or assuming statistical independence among records. Large entanglement in the perturbed branch becomes locally informative precisely because the unperturbed branch remains confined to a recognizable low-entropy support.

Quantitatively, let a macroscopic interval $A$ contain $m=\Theta(L)$ complete cells and satisfy $S(\rho_{1,A})\ge\sigma m$ with a fixed entropy density $\sigma>0$ in bits per cell. Combining entropy subadditivity with the concavity of $g_D(q)$ yields
\begin{equation}
 \sigma\le\frac{S(\rho_{1,A})}{m}
 \le\frac1m\sum_a g_D(q_a)\le g_D(\bar q),
 \label{eq:entropy}
\end{equation}
where $\bar q=m^{-1}\sum_aq_a$ is the mean departure probability. Let $q_\sigma\in(0,1-1/D]$ denote the unique root of $g_D(q_\sigma)=\sigma$, so that $\bar q\ge q_\sigma$. Because every $q_a\le1$, Markov's inequality implies that at least a fraction $q_\sigma/(2-q_\sigma)$ of the cells must satisfy $q_a\ge q_\sigma/2$. Thus, the \emph{product-reference case} yields the explicit positive threshold $\eps_{\rm prod}=f(q_\sigma/2)>0$ and the extensive record count $R_{\eps_{\rm prod}}(A,t)\ge mq_\sigma/(2-q_\sigma)$.

This conclusion generalizes directly to any \emph{uniform area-law reference} upon replacing the single-cell projector with an approximate reference-subspace projector on blocks of a sufficiently large yet fixed length $\ell$. The area law bounds the rank of this dominant subspace independently of $L$ for any chosen leakage tolerance $\eta>0$. Volume-law entanglement in the perturbed branch then forces a finite fraction of $\ell$-site blocks to depart from their approximate reference supports with probability exceeding $\eta$, ensuring that the corresponding binary tests distinguish the two branches. Under the hypotheses of our main result, we obtain
\begin{equation}
 R_{\eps_*}(L,t_L)\ge\kappa L,\qquad
        \eps_*,\kappa>0,\quad \ell<\infty,
 \label{eq:threshold}
\end{equation}
where all constants are independent of system size. The complete proof in Appendix~\ref{app:entropy} rigorously incorporates finite reference leakage and provides an explicit lower bound on the record density $\kappa$. Because the derivation relies solely on the conditional branch entropies at the observation time, it accommodates arbitrary correlations between distinct fragments.

Our criterion cleanly separates kinematic state structure from microscopic dynamics. Whenever the stated entanglement contrast is achieved, extensive partial records must exist; local thermalization of the perturbed branch is one natural physical mechanism that generates this contrast. Determining which local Hamiltonians dynamically produce such an entanglement contrast is a separate question, which we now address through explicit many-body realizations using simple, fixed local readouts.

\section{Dynamical selection in stationary and recurrent references}
\label{sec:models}
\subsection{Embedded dimers with local activation and reliable records}
\label{sec:dimer}
On an even periodic spin-$1/2$ chain of length $L$, we prepare the reference singlet covering $\ket\Omega=\bigotimes_{k=0}^{L/2-1}\ket{s}_{2k,2k+1}$, where $\ket s=(\ket{\uparrow\downarrow}-\ket{\downarrow\uparrow})/\sqrt2$. Applying the local Pauli operator $Z_0$ converts the $k=0$ singlet into its zero-magnetization triplet. At $t=0$, only that single dimer distinguishes the two branches, and all single-spin marginals are maximally mixed in both preparations.

We evolve the system under a Shiraishi--Mori embedded Hamiltonian~\cite{Shiraishi2017},
\begin{equation}
 H_D=\sum_iP_i^{(3/2)}h_iP_i^{(3/2)},\qquad H_D\ket\Omega=0,
 \label{eq:dimerH}
\end{equation}
where $P_i^{(3/2)}$ projects the three consecutive spins $\{i-1,i,i+1\}$ onto their total-spin-$3/2$ subspace. Because every three-spin cluster in $\ket\Omega$ contains a complete two-spin singlet and thus has total spin $1/2$, $P_i^{(3/2)}$ annihilates $\ket\Omega$ identically. The local operators $h_i$ and fixed coupling constants are specified in Appendix~\ref{app:dimer}. The seeded state $Z_0\ket\Omega$ has a conserved mean energy of $1.2J$ and an energy width of order $J$, ensuring that both branches share the exact same thermodynamic energy density.

The projector structure also governs how the local defect spreads. Defining the singlet projector $P_{s,k}=\ket{s_k}\bra{s_k}$ and triplet projector $D_k=1-P_{s,k}$ on reference dimer $k$, the two Hamiltonian terms anchored on dimer $k$ combine into $G_k=D_kG_kD_k$, which acts nontrivially only on dimer $k$ and its immediate neighbors $k\pm1$. Triplet weight on dimer $k$ therefore facilitates singlet--triplet transitions on adjacent bonds, represented schematically as $TS\rightleftarrows TT$ and $ST\rightleftarrows TT$. Although coherent quantum evolution permits destructive interference and local singlet recovery, this kinetic facilitation allows finite triplet probability to proliferate across many bonds while the unperturbed reference preserves every singlet intact.

Crucially, this proliferation differs fundamentally from the transport of a single conserved quasiparticle. Against a vacuum reference, a single conserved excitation obeys $\sum_aq_a=1$ and can sustain at most $1/f^{-1}(\eps)$ occupation readers above threshold $\eps$, regardless of how far it propagates. In our embedded-dimer chain, triplet number is not conserved, and because the local operators $h_i$ have both positive and negative eigenvalues, creating additional triplets does not incur a fixed positive energy penalty. Consequently, the fixed binary test $\{P_{s,k},1-P_{s,k}\}$ can witness triplet activity throughout an expanding spatial region.

\begin{figure}[!t]
\centering
\includegraphics[width=\columnwidth]{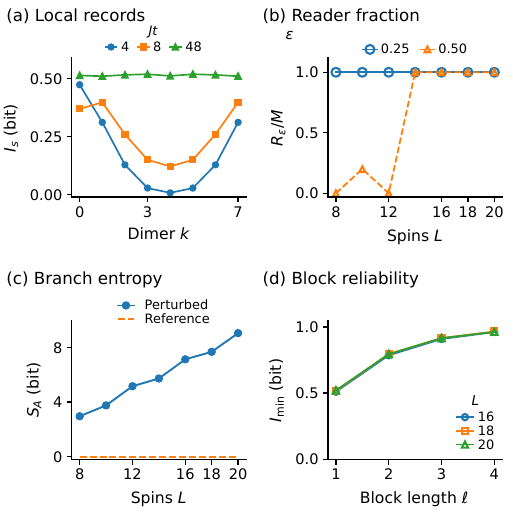}
\caption{\label{fig:dimer}Information replication from a stationary singlet reference. (a) Single-dimer accessible information for $L=16$ at $Jt=4,8,48$, showing how the initially localized imprint becomes spatially distributed. (b) Fraction of simultaneous single-dimer readers at $Jt=3L$ across $M=L/2$ disjoint dimers. Every reference dimer exceeds $0.50$ bits for $L=14,16,18,20$. (c) Entanglement entropies $S_1$ and $S_2$ of the perturbed branch on an interval of $\lfloor M/2\rfloor$ complete dimers starting at $k=0$, across which the reference entropy vanishes identically. The interval comprises eight spins at $L=16,18$ and ten spins at $L=20$. (d) Minimum accessible information across all cyclic positions of an $\ell$-dimer block at $Jt=3L$, where four-dimer records surpass $0.96$ bits for all three chain lengths. Solid lines connect finite-ring data rather than asymptotic scaling fits (see Appendix~\ref{app:records} for detailed definitions and numerical values).}
\end{figure}

Figure~\ref{fig:dimer}(a) illustrates how the initial local imprint evolves into comparable readable information across distant bonds. Evaluated at the scaled clock $Jt=3L$, every reference dimer surpasses $0.25$ bits across all examined even chain lengths $L=8$--$20$, and every dimer also exceeds $0.50$ bits for $L=14,16,18,20$ [Fig.~\ref{fig:dimer}(b)]. All system sizes employ the exact same two-outcome singlet test on each bond.

While the reference branch remains unentangled across dimer-aligned cuts, the perturbed branch develops substantial bipartite entanglement [Fig.~\ref{fig:dimer}(c)]. At $L=20$, a five-dimer half-chain carries $S_1\simeq9.06$ bits of von Neumann entropy. Substituting this value into our product-reference bound with $D=4$ certifies at least two $0.14$-bit records within that half-chain from entanglement alone, whereas direct numerical readout shows that all ten dimers on the ring exceed $0.52$ bits. Physically, the entropy bound reflects the fact that each dimer has a single singlet state but three orthogonal triplet states; accommodating large entanglement forces local marginals to populate the triplet sector without requiring any preferred spin polarization, which is why a simple polarization-blind singlet test reliably detects the branch contrast (see Appendix~\ref{app:records}).

Grouping adjacent dimers into larger observation blocks substantially enhances readout reliability. Suppose an observer tests a contiguous block $F$ of $\ell$ reference dimers, inferring the perturbed branch if any triplet is detected and inferring the reference branch only if all $\ell$ bonds project onto singlets. Letting $w_\ell(F,t)=\langle\prod_{k\in F}P_{s,k}\rangle_1$ denote the joint all-singlet probability in the perturbed branch, the equal-prior error probability and accessible block information are given by
\begin{equation}
 P_{\rm err}=\frac{w_\ell}{2},\qquad I_\ell=f(1-w_\ell),
 \label{eq:reliability}
\end{equation}
because false-positive errors are absent on the exact singlet reference. If each relaxed dimer independently yielded a singlet with probability $p_s=1/4$, one would have $w_\ell=p_s^\ell$ and $w_4=1/256$. Our exact numerical calculations evaluate the correlated joint probability $w_\ell(F,t)$ directly from the full many-body state.

As shown in Fig.~\ref{fig:dimer}(d), even the least informative four-dimer block on the ring exceeds $0.96$ bits at $L=16,18,20$, guaranteeing at least two disjoint, highly reliable records at the common clock $Jt=3L$. Because Fig.~\ref{fig:dimer}(d) reports the minimum over all cyclic block translations, high readout reliability holds throughout the chain rather than only near the seed. Equilibrium large-deviation theory explains why the all-singlet probability decays exponentially with block length $\ell$ even in the presence of short-range spatial correlations~\cite{Araki1969,Ogata2010,Zwolak2014}, and Appendix~\ref{app:records} compares our dynamical block probabilities against energy-matched canonical Gibbs benchmarks.

Numerically, we employ exact sparse matrix exponentiation for $L\le16$ and matrix-free Krylov evolution for $L=18,20$. While these results establish simultaneous partial records and multi-dimer reliability on finite rings, we note that the clock $Jt=3L$ allows perturbations to wrap around the periodic boundary and that four-dimer blocks span $40$--$50\%$ of these chain lengths. Proving the thermodynamic persistence of the dynamical entropy density remains an open problem distinct from the kinematic sufficiency criterion of Sec.~\ref{sec:theory}.

\subsection{Records against a coherently recurrent reference in the PXP model}
\label{sec:pxp}
Information replication can also operate against a time-dependent reference. In the Rydberg-blockaded PXP chain, the staggered N\'eel state $\ket{Z_2}$ undergoes coherent scar revivals~\cite{Turner2018,TurnerPRB2018,Ho2019,Lin2019}, providing a recognizable local spin pattern at discrete revival phases. We compare the standard and stabilized periodic PXP Hamiltonians~\cite{Choi2019},
\begin{equation}
 \frac{H_\lambda}{J}=\sum_jP_{j-1}X_jP_{j+1}
 [1-\lambda(Z_{j-2}+Z_{j+2})],
 \label{eq:pxp}
\end{equation}
where $P_j=(1-Z_j)/2$ and $\lambda=0$ or $0.053$. The input bit selects either $\ket{Z_2}$ or the single-defect preparation $X_{j_0}\ket{Z_2}$ obtained by de-exciting one site, both of which have zero mean energy. Measuring individual spins in the laboratory $Z$ basis yields conditional outcome probabilities $p_{b,j}^{\pm}=(1\pm\langle Z_j\rangle_b)/2$ and single-site information $I_Z(j,t)$ via Eq.~\eqref{eq:mutual}.

\begin{figure}[!t]
\centering
\includegraphics[width=\columnwidth]{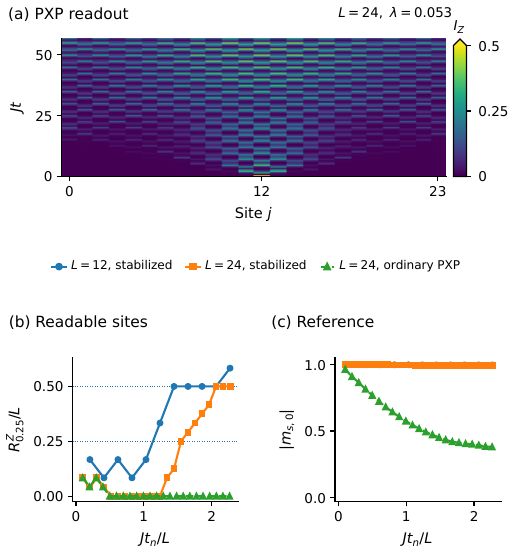}
\caption{\label{fig:pxp}Information replication against a coherently recurrent reference. (a) Space-time profile of single-site fixed-$Z$ information for stabilized PXP ($\lambda=0.053$) at $L=24$, with the color scale saturated at $0.50$ bits. (b) Fraction of $0.25$-bit readers evaluated at successive reference revival peaks. Stabilized PXP reaches twelve simultaneous readers at $L=24$, whereas ordinary PXP ($\lambda=0$) loses all quarter-bit records at late revivals. (c) Reference staggered-magnetization amplitude $|m_{s,0}|$ at the corresponding revival events. Stabilized PXP is shown for $L=12,24$, and ordinary PXP at $L=24$ is evaluated at its own reference peaks. All sites and both conditional branches share a single observation time at each revival event, with lines connecting discrete stroboscopic readouts (see Appendix~\ref{app:pxp} for event selection and basis-optimization checks).}
\end{figure}

The space-time profile in Fig.~\ref{fig:pxp}(a) exhibits an expanding light cone of readable contrast modulated by the underlying scar oscillations. For each coupling $\lambda$, we select stroboscopic observation times purely from the local maxima of the unperturbed reference staggered magnetization $|m_{s,0}(t)|$, where $m_{s,0}=L^{-1}\sum_j(-1)^j\langle Z_j\rangle_0$. Evaluating all sites at a single common revival time is essential when the reference oscillates, as optimizing readout times separately at each site would conflate measurements from different physical times.

Our exact constrained-subspace simulations span $L=12,16,20,24$. At $L=24$, stabilized PXP first achieves six quarter-bit readers near $Jt=37.2$ and twelve readers near $Jt=49.6$, with high record counts recurring across subsequent revival peaks [Fig.~\ref{fig:pxp}(b)]. By contrast, ordinary PXP ($\lambda=0$) yields zero quarter-bit readers from its seventh revival peak onward as its reference contrast progressively dephases [Fig.~\ref{fig:pxp}(c)]. Quasi-exact scar stabilization thus preserves a sharp reference pattern long enough for the local perturbation to spread across the chain.

This late-revival contrast is robust against local measurement optimization. At $L=24$, the final three sampled revivals of stabilized PXP each support twelve $0.25$-bit records under both fixed-$Z$ measurements and site-optimized projective readouts, whereas ordinary PXP supports none from its seventh revival onward even under the Holevo upper bound (Appendix~\ref{app:pxp}). At the final stabilized revival, all $24$ sites surpass $0.10$ bits while $12$ sites surpass $0.25$ bits, illustrating the trade-off between record multiplicity and the information threshold demanded by each observer.

Both differential mixedness and Bloch-vector misalignment contribute to single-site distinguishability in PXP. At late revival peaks, the stabilized reference returns near pure, locally polarized states, whereas the perturbed branch remains locally mixed with rotated Bloch vectors. While our entropy criterion captures the distinguishability arising from local mixedness, coherent spin rotation provides an additional resource. Rydberg-atom quantum simulators, which have already demonstrated scar revivals and Holevo-information dynamics, offer a direct experimental platform to test these stroboscopic local readouts~\cite{Bernien2017,Yuan2022,Xiang2026,Chen2026}.


\section{Conclusions and outlook}
\label{sec:conclusions}
We have introduced information replication to characterize the extensive spatial multiplicity of accessible local records generated from a single locally encoded classical bit. By identifying dynamical selection as an underlying physical mechanism, we proved a general entropy-capacity criterion showing that an area-law reference paired with a volume-law perturbed branch guarantees extensive records above a fixed positive threshold. Our numerical investigations of embedded dimers and stabilized PXP chains demonstrate the formation of simultaneous local records against both stationary and coherently recurrent references in finite systems.

Distinguishing between causal propagation and the local retention of readable evidence opens several theoretical avenues. Other nonthermal eigenstates, coherent amplification protocols, and kinetically constrained models may furnish recognizable references with distinct lifetimes and readout observables. A central open dynamical question is to determine the precise conditions under which a single local imprint generates a macroscopic entanglement contrast. While Lieb--Robinson locality sets a linear lower bound on the propagation time in one dimension, local equilibration can proceed more slowly, leaving the asymptotic thermodynamic record density and long-time persistence as important targets for future work.

Finally, the proliferation of local records connects naturally to quantum Darwinism, where disjoint environment fragments redundantly encode information about a system to establish classical objectivity~\cite{Ollivier2004,Ollivier2005,Zurek2009,Girolami2022,Doucet2024,Ferte2024}. In our framework, the classical bit is encoded via a single local operation and replicated entirely within a closed many-body system without assuming an external system--environment partition. Because our criterion addresses locally accessible classical information rather than full spectrum broadcast structure~\cite{Le2019}, clarifying how inter-fragment correlations and sequential measurement backaction affect multi-observer consensus represents a natural next step. Experimentally, witnessing replication requires only two initial preparations and fixed local readout statistics, enabling direct tests of record multiplicity and block reliability without full state tomography.

\begin{acknowledgments}
    
This work was supported by the National Natural Science Foundation of China (Grant No.12375013, No.12547109), the Guangdong Basic and Applied Basic Research Fund (Grant No.2023A1515011460), and Guangdong Provincial Quantum Science Strategic Initiative (Grant No. GDZX2503008).
    
\end{acknowledgments}

\label{maintextend}

\appendix
\section{Readout information and record budgets}
\label{app:information}
This appendix establishes the operational conventions used in Sec.~\ref{sec:theory} and derives the record-counting bounds for binary reference tests. Throughout this work, the initial bit is equiprobable, all entropies and mutual information quantities use base-two logarithms, and all observers measure their disjoint fragments at a single common observation time. Repeated experimental runs serve solely to sample the conditional outcome distributions rather than to supply additional independent input bits.

For the equal-prior local ensemble $\{\tfrac12,\rho_{0,F};\tfrac12,\rho_{1,F}\}$, the Holevo quantity is defined as
\begin{equation}
 \chi_F=S\!\left(\frac{\rho_{0,F}+\rho_{1,F}}2\right)
 -\frac{S(\rho_{0,F})+S(\rho_{1,F})}{2}.
 \label{eq:app-holevo}
\end{equation}
Any specific measurement $\mathcal M$ obeys the hierarchy $I_{\mathcal M}\le I_{\rm acc}\le\chi_F\le1$, where $I_{\rm acc}$ maximizes the mutual information over all positive operator-valued measures (POVMs)~\cite{Holevo1973,Zwolak2013}. While $\chi_F$ provides a useful measurement-independent upper bound, all record counts in the main text are evaluated directly from the specified physical readouts.

For an ideal binary reference test $\{P,1-P\}$ satisfying $\Tr(P\rho_0)=1$ and $\Tr[(1-P)\rho_1]=q$, the conditional outcome distributions are $(1,0)$ and $(1-q,q)$. Substituting these into Eq.~\eqref{eq:mutual} yields $I=f(q)$ as defined in Eq.~\eqref{eq:f}, whose derivative is strictly positive for $q\in(0,1)$,
\begin{equation}
 f'(q)=\frac12\log_2\frac{2-q}{1-q}>0.
 \label{eq:app-fprime}
\end{equation}
Consequently, a cell is readable at threshold $\eps$ if and only if its departure probability satisfies $q\ge q_\eps=f^{-1}(\eps)$. Given $M$ elementary cells with mean departure probability $\nu=M^{-1}\sum_aq_a$, partitioning the sum into readable and unreadable cells yields the two-sided bound
\begin{equation}
 M\left[\frac{\nu-q_\eps}{1-q_\eps}\right]_+
 \le R_\eps^{(1)}\le\frac{M\nu}{q_\eps},
 \qquad [x]_+=\max(x,0),
 \label{eq:app-count}
\end{equation}
where the lower bound follows from $M\nu=\sum_aq_a\le R_\eps^{(1)}+(M-R_\eps^{(1)})q_\eps$ and the upper bound follows from $M\nu\ge R_\eps^{(1)}q_\eps$. Thus, a strictly positive asymptotic departure density $\nu$ guarantees extensive single-cell records below a corresponding fixed threshold. These bounds hold for arbitrary spatial profiles and inter-cell correlations, and non-integer lower bounds can be rounded up to the next integer at any fixed system size.

When the reference state is a pure product across elementary cells with projectors $P_a=\ket{\phi_a}\bra{\phi_a}$ and $D_a=1-P_a$, a multi-cell fragment $F$ has the joint reference projector $P_F=\prod_{a\in F}P_a$. Because projectors on disjoint cells commute, the operator union bound gives
\begin{equation}
 1-P_F\le\sum_{a\in F}D_a,\qquad
 q_F\le\sum_{a\in F}q_a.
 \label{eq:app-union}
\end{equation}
Summing Eq.~\eqref{eq:app-union} over any collection of mutually disjoint readable fragments yields the global budget $R_\eps^{\rm ref}\le\sum_aq_a/q_\eps$. In particular, a single conserved excitation ($\sum_aq_a=1$) can support at most $1/q_\eps$ readers, as noted in Sec.~\ref{sec:dimer}. More generally, if a conserved operator $\mathcal C\ge\Delta\sum_aD_a$ with $\Delta>0$ has a bounded expectation value $E=\langle\mathcal C\rangle_1=O(1)$, then $R_\eps^{\rm ref}\le E/(\Delta q_\eps)$ for any disjoint family of complete-cell reference tests.

When evaluating scaling across system sizes, the fragment family, threshold $\eps$, and observation protocol are held fixed. While $R_\eps$ denotes the maximum over all disjoint members of that family, any specific disjoint partition provides a rigorous lower bound. Varying the block length $\ell$ at fixed $L$ probes single-record reliability, whereas holding $\ell$ and $\eps$ fixed as $L$ grows probes spatial extensivity. Similarly, evaluating counts at discrete revival peaks tests stroboscopic recurrence rather than continuous-time storage between revivals.

\section{Proof of the entanglement-contrast criterion}
\label{app:entropy}
This appendix proves the main theoretical result of Sec.~\ref{sec:theory}, including the generalization to approximate reference supports. The proof is a kinematic statement relating the entropies of two conditional pure states at their common readout time, requiring neither canonical thermalization nor dynamical assumptions on equilibration timescales.

\subsection{Entropy capacity and readable blocks}
Let $P$ project onto a rank-$r$ subspace of a $D$-dimensional block and define the departure weight $q=\Tr[(1-P)\rho]$. The projective pinching map $\rho\mapsto P\rho P+(1-P)\rho(1-P)$ is an equal mixture of the unitary conjugations $\rho$ and $(2P-1)\rho(2P-1)$, and therefore cannot decrease the von Neumann entropy. Because the two pinched components reside in orthogonal subspaces of dimensions $r$ and $D-r$, respectively, we obtain
\begin{align}
 S(\rho)&\le h_2(q)+(1-q)S\!\left(\frac{P\rho P}{1-q}\right)\nonumber\\
 &\quad+qS\!\left(\frac{(1-P)\rho(1-P)}q\right)\nonumber\\
 &\le g_{D,r}(q),
 \label{eq:app-pinching}\\
 g_{D,r}(q)&=h_2(q)+(1-q)\log_2r+q\log_2(D-r),
 \label{eq:app-capacity}
\end{align}
where boundary values at $q=0,1$ are defined by continuity. Differentiating gives $g'_{D,r}(q)=\log_2[(1-q)(D-r)/(qr)]$, confirming that $g_{D,r}(q)$ increases monotonically on $0\le q\le 1-r/D$.

For an exact rank-$r$ reference support ($\Tr(P\rho_0)=1$), the binary test extracts $f(q)$ bits of information. Fixing a threshold $\eps$ such that $q_\eps=f^{-1}(\eps)<1-r/D$, any unreadable block has $q<q_\eps$ and thus carries at most $b_\eps=g_{D,r}(q_\eps)$ bits of entropy, whereas each readable block carries at most $\log_2D$ bits. Partitioning an interval $A$ into $m$ disjoint blocks and applying entropy subadditivity yields
\begin{align}
 S(\rho_{1,A})&\le(m-R)b_\eps+R\log_2D,\nonumber\\
 R_\eps(A,t)&\ge
 \left[\frac{S(\rho_{1,A})-mb_\eps}{\log_2D-b_\eps}\right]_+.
 \label{eq:app-capacity-count}
\end{align}
This bound holds regardless of spatial correlations among the blocks, relying solely on their mutual disjointness and individual entropy ceilings.

For a pure product reference ($r=1$), Eq.~\eqref{eq:app-capacity} reduces to $g_D(q)$ in Eq.~\eqref{eq:capacity}. Concavity of $g_D(q)$ and entropy subadditivity then give Eq.~\eqref{eq:entropy}. Because the smaller root $q_\sigma$ of $g_D(q)=\sigma$ is strictly positive for $0<\sigma\le\log_2D$, the inequality $g_D(\bar q)\ge\sigma$ forces $\bar q\ge q_\sigma$. Applying the lower bound in Eq.~\eqref{eq:app-count} at the departure threshold $q_\sigma/2$ immediately establishes the explicit product-reference estimate in Sec.~\ref{sec:theory}.

\subsection{Formal statement and finite-leakage proof}
\noindent\textbf{Proposition.}
Consider conditional pure states $\ket{\psi_b(t_L)}$ on chains of local dimension $d$ evaluated at a common sequence of times $t_L$. Suppose that, uniformly for all sufficiently large $L$,
\begin{align}
 S(\rho_{0,F}(t_L))&\le C\quad\text{for every contiguous interval }F,\nonumber\\
 S(\rho_{1,A_L}(t_L))&\ge s|A_L|,\qquad |A_L|\ge aL,
 \label{eq:app-hypotheses}
\end{align}
where $C<\infty$ and $s,a>0$ are constants independent of $L$. Then there exist system-size-independent constants $\ell,\eps_*,\kappa>0$ and binary projective tests on disjoint contiguous $\ell$-site blocks such that $R_{\eps_*}(L,t_L)\ge\kappa L$ for all sufficiently large $L$. Each local projector may depend on the block position and observation time $t_L$, but is independent of the unknown input bit $b$.

To prove this proposition, suppose a block projector $P$ has reference leakage $u=\Tr[(1-P)\rho_0]\le\eta$. The binary test $\{P,1-P\}$ yields mutual information
\begin{equation}
 I(u,q)=h_2\!\left(\frac{u+q}{2}\right)
 -\tfrac12h_2(u)-\tfrac12h_2(q).
 \label{eq:app-leakage}
\end{equation}
In the regime $q>u$, $I(u,q)$ increases monotonically with $q$ and decreases monotonically with $u$. Selecting fixed constants $0<\eta<q_*<s/\log_2d$, any block with $q\ge q_*$ achieves at least $I(u,q)\ge I(\eta,q_*)\equiv\eps_*>0$ bits. Consequently, any block that fails to reach the threshold $\eps_*$ must satisfy $q<q_*$.

The uniform area law on the reference branch guarantees a low-rank projector satisfying $u\le\eta$. Treating the eigenvalues $\{\lambda_i\}$ of $\rho_{0,F}$ as probabilities, the expectation value of the nonnegative surprisal $-\log_2\lambda_i$ is $S(\rho_{0,F})\le C$. By Markov's inequality, the total weight of eigenvalues with $-\log_2\lambda_i>C/\eta$ (i.e., $\lambda_i<2^{-C/\eta}$) is at most $\eta$. Projecting onto the eigenspace with $\lambda_i\ge2^{-C/\eta}$ therefore captures weight at least $1-\eta$ and has rank at most $r_0=\lceil2^{C/\eta}\rceil$, independent of both $|F|$ and $L$ (with $r_0=1$ when $C=0$). Once the block dimension exceeds $r_0$, any lower-rank support can be padded arbitrarily to rank $r_0$.

For contiguous blocks of $\ell$ sites, the Hilbert-space dimension is $D=d^\ell$. Because $g_{d^\ell,r_0}(q_*)/\ell\to q_*\log_2d<s$ as $\ell\to\infty$, we can fix a finite block length $\ell$ large enough that $q_*<1-r_0/d^\ell$ and
\begin{equation}
 b_*:=g_{d^\ell,r_0}(q_*)<s\ell.
 \label{eq:app-blockchoice}
\end{equation}
Partitioning the macroscopic interval $A_L$ into $m_L=\lfloor|A_L|/\ell\rfloor$ complete $\ell$-site blocks leaves a boundary remainder of fewer than $\ell$ sites, whose entropy cannot exceed $\ell\log_2d$. Applying Eq.~\eqref{eq:app-capacity-count} then gives the asymptotic fraction of readable blocks
\begin{equation}
 \liminf_{L\to\infty}\frac{R_{\eps_*}(L,t_L)}{m_L}
 \ge\frac{s\ell-b_*}{\ell\log_2d-b_*}=:\zeta>0.
 \label{eq:app-density}
\end{equation}
Because $m_L\ge aL/\ell-O(1)$, any positive constant $\kappa<a\zeta/\ell$ satisfies $R_{\eps_*}(L,t_L)\ge\kappa L$ for all sufficiently large $L$. Notice that the constants $\ell,\eps_*,\kappa$ are fixed prior to taking the large-$L$ limit. In particular, if the entropy hypotheses hold at the linear clock $Jt_L=cL+O(1)$, extensive replication is achieved on that timescale.

\section{Embedded-dimer Hamiltonian and numerical evolution}
\label{app:dimer}
This appendix details the embedded-dimer model analyzed in Fig.~\ref{fig:dimer} and derives its exact reference and local activation properties. We consider periodic chains of even length $L$ containing $M=L/2$ reference dimers. Writing the spin-$1/2$ operators as $\mathbf S_j=\boldsymbol\sigma_j/2$, the Hamiltonian reads
\begin{align}
 H_D&=\sum_iH_i,\qquad H_i=P_i^{(3/2)}h_iP_i^{(3/2)},\nonumber\\
 P_i^{(3/2)}&=\frac12+\frac23
 \sum_{a<b\in\{i-1,i,i+1\}}\mathbf S_a\cdot\mathbf S_b,
 \label{eq:app-dimerH}\\
 h_i&=\sum_{\mu=x,y,z}\big[J_\mu S_i^\mu(S_{i-1}^\mu+S_{i+1}^\mu)\nonumber\\
 &\hspace{14mm}+J'_\mu S_i^\mu(S_{i-2}^\mu+S_{i+2}^\mu)-h_\mu S_i^\mu\big].
 \label{eq:app-dimerlocal}
\end{align}
Throughout all calculations, we fix the couplings to $\mathbf J/J=(1,1,-0.6)$, $\mathbf J'/J=(-0.8,0,0)$, and $\mathbf h/J=(0.3,0,0.1)$, so each dressed local term $H_i$ acts on at most five consecutive spins.

The unperturbed reference state is $\ket\Omega=\bigotimes_{k=0}^{M-1}\ket{s}_{2k,2k+1}$, and the seeded state is $\ket\phi=Z_0\ket\Omega$. Because every three-spin cluster $\{i-1,i,i+1\}$ in either of the two translationally related singlet coverings contains an intact two-spin singlet, its total spin is $1/2$, so $P_i^{(3/2)}$ annihilates both coverings identically~\cite{Shiraishi2017} and $H_D\ket\Omega=0$. The seeded state $\ket\phi$ is orthogonal to $\ket\Omega$ because $Z_0$ converts the $k=0$ singlet into its $S^z=0$ triplet, and it is also orthogonal to the shifted singlet covering because both coverings are global spin singlets ($S_{\rm tot}=0$) whereas the vector operator $Z_0$ has vanishing matrix elements between any two total-spin singlets. This kinematic orthogonality is a property of the states alone and does not require $H_D$ to be $\mathrm{SU}(2)$ invariant.

Defining the triplet projector $D_k=1-P_{s,k}$ on reference dimer $k$ and grouping the two terms anchored on dimer $k$ as $G_k=H_{2k}+H_{2k+1}$, the fact that $P_{2k}^{(3/2)}$ and $P_{2k+1}^{(3/2)}$ annihilate $\ket{s_k}$ implies the operator identity
\begin{equation}
 G_k=D_kG_kD_k,\qquad
 \operatorname{supp}G_k\subset\{k-1,k,k+1\}
 \label{eq:app-anchor}
\end{equation}
in terms of the dimer index $k$. Thus, a singlet anchor is completely inert under $G_k$, whereas triplet weight on dimer $k$ enables transitions on the adjacent dimers $k\pm1$. This exact operator identity underlies the reversible kinetic facilitation described in Sec.~\ref{sec:dimer}.

For all $L\ge6$, the seeded state $\ket\phi$ has a conserved mean energy of $1.2J$ and an energy variance of approximately $0.31J^2$. More generally, applying any local operator $V_{j_0}$ to an exact eigenstate $\ket\Omega$ gives
\begin{equation}
 (H-E_0)V_{j_0}\ket\Omega=[H,V_{j_0}]\ket\Omega,
 \label{eq:app-seed-energy}
\end{equation}
which involves only Hamiltonian terms overlapping the bounded support of $V_{j_0}$ and thus imparts an energy width of order $J$. For our coupling parameters, the infinite-temperature state has an energy density of $7J/60$ per spin, while the fully $+z$-polarized state has an energy density of $-7J/20$ per spin. Because the zero energy density of $\ket\Omega$ lies strictly within the interior of the spectrum, energy conservation does not impose a positive energy penalty per triplet to obstruct defect proliferation.

Numerically, states for $L\le16$ are evolved via exact sparse matrix exponentiation, while for $L=18,20$ we employ a matrix-free Krylov subspace propagator expressed in the four-level singlet--triplet dimer basis without storing the global sparse matrix. The Krylov implementation was verified against the spin-basis evolution at $L=16$ to within $10^{-14}$, with state norms preserved to the same precision across $L=16,18,20$. No tensor-network truncation is used.

In Fig.~\ref{fig:dimer}(c), the subsystem cut begins at dimer $k=0$ and spans $n_A=\lfloor M/2\rfloor$ complete reference dimers. The entropies $S_1=-\sum_\alpha p_\alpha\log_2p_\alpha$ and $S_2=-\log_2\sum_\alpha p_\alpha^2$ are extracted from the exact Schmidt probabilities $p_\alpha$ and cross-checked against direct diagonalization of the reduced density matrix. Rather than averaging spatially, Fig.~\ref{fig:dimer}(c) reports the entropy for the fixed interval starting at $k=0$, while all cyclic translations are retained for the block-readout analysis.

\section{Dimer record statistics and block reliability}
\label{app:records}
This appendix provides the numerical values supporting Fig.~\ref{fig:dimer} and formulates the local-relaxation conditions required to connect equilibrium thermal statistics to dynamical block reliability. All measurements refer to the original dimer covering under equal branch priors.

\subsection{Simultaneous readers and entropy witnesses}
For single-dimer readouts, the accessible information is $I_k=f(q_k)$ with $q_k=1-\langle P_{s,k}\rangle_1$, and reader counts include the boundary case $I_k=\eps$. At the scaled clock $Jt=3L$, the $0.10$- and $0.25$-bit reader counts equal the maximum $M=L/2$ for all examined even chain lengths $L=8$--$20$, and all $M$ dimers also surpass $0.50$ bits for $L=14,16,18,20$ (Table~\ref{tab:dimer}).

\begin{table}[!htbp]
\caption{\label{tab:dimer}Embedded-dimer benchmarks at $Jt=3L$. Here $R_{0.50}$ denotes the number of $0.50$-bit readers among the $M=L/2$ reference dimers, while $S_1$ and $S_2$ denote the entropies of the interval starting at $k=0$ containing $n_A=\lfloor M/2\rfloor$ complete dimers. All entropies and information values are expressed in bits.}
\centering
\begin{ruledtabular}
\begin{tabular}{rrrrrr}
$L$ & $n_A$ & $\min_k I_k$ & $R_{0.50}$ & $S_1$ & $S_2$\\
16 & 4 & 0.51029 & 8 & 7.13533 & 6.81946\\
18 & 4 & 0.51752 & 9 & 7.68478 & 7.48676\\
20 & 5 & 0.52091 & 10 & 9.05505 & 8.74283
\end{tabular}
\end{ruledtabular}
\end{table}

Note that preserving dimer-aligned cuts is essential in Fig.~\ref{fig:dimer}(c). Both $L=16$ and $L=18$ use an eight-spin subsystem ($n_A=4$), whereas $L=20$ uses a ten-spin subsystem ($n_A=5$). Taking a nine-spin half-chain at $L=18$ would sever a reference singlet and artificially introduce one bit of reference entropy; restricting cuts to complete dimers keeps the reference entropy identically zero.

To evaluate the product-reference witness at $L=20$, taking $\sigma=9.05505/5$ bits per dimer yields $q_\sigma\simeq0.512$ from $h_2(q_\sigma)+q_\sigma\log_2 3=\sigma$. Substituting this into our product bound gives $\eps_{\rm prod}=f(q_\sigma/2)\simeq0.14159$ bits and $R_{\eps_{\rm prod}}\ge1.720$, certifying at least two $0.14$-bit readers within that five-dimer interval from entropy alone. Similarly, at $L=12$ and $Jt=36$, each three-dimer half-chain has $S_1\simeq5.16$ bits; for $\eps=0.10$ bits, Eq.~\eqref{eq:app-capacity-count} gives $b_\eps\simeq0.98853$ bits and $R_{0.10}>2$ in each half, certifying all six dimers on the ring as simultaneous readers.

\subsection{Joint block probabilities and the thermal comparison}
For a contiguous block $F$ of $\ell$ reference dimers, the joint all-singlet projector is $P_F=\prod_{k\in F}P_{s,k}$ with expectation value $w_\ell=\Tr(P_F\rho_{1,F})$. Assigning any triplet detection to branch $1$ and the all-singlet outcome to branch $0$ yields the error probability and mutual information in Eq.~\eqref{eq:reliability} without assuming statistical independence between dimers.

\begin{table}[!htbp]
\caption{\label{tab:blocks}Minimum accessible block information $I_{\min}$ over all cyclic block positions at $Jt=3L$ for blocks of $\ell$ reference dimers. The final two columns compare the spatially averaged four-dimer all-singlet probability $\bar w_4$ in the exact time-evolved state against the energy-matched canonical Gibbs ensemble.}
\centering
\begin{ruledtabular}
\begin{tabular}{rrrrrr}
 & \multicolumn{3}{c}{$I_{\min}$ (bit)} & \multicolumn{2}{c}{$\bar w_4$}\\
$L$ & $\ell=2$ & $\ell=3$ & $\ell=4$ & dynamics & Gibbs\\
16 & 0.78526 & 0.90855 & 0.96047 & 0.009156 & 0.004374\\
18 & 0.79470 & 0.91557 & 0.96433 & 0.008268 & 0.004443\\
20 & 0.79454 & 0.91503 & 0.96269 & 0.008369 & 0.004506
\end{tabular}
\end{ruledtabular}
\end{table}

Table~\ref{tab:blocks} confirms that the worst-case four-dimer block on the ring exceeds $0.96$ bits across $L=16,18,20$, guaranteeing at least two disjoint, highly reliable records at each size. Because four-dimer blocks span $40$--$50\%$ of these finite chains, these results demonstrate finite-system block reliability rather than an asymptotic thermodynamic limit at fixed $\ell$.

In a one-dimensional finite-range Gibbs state $\omega_\beta$ at finite inverse temperature $\beta$ with a positive triplet density $\rho_T^\beta>0$, the all-singlet event becomes exponentially rare in $\ell$ despite spatial correlations. Defining the block triplet number operator $N_F=\sum_{k\in F}(1-P_{s,k})$ so that $P_F=\mathbf1_{N_F=0}$, Markov's inequality for any $u>0$ gives
\begin{equation}
 w_\ell^\beta\le\omega_\beta(e^{-uN_F}),\qquad
 \phi_\beta(z)=\lim_{\ell\to\infty}\frac1\ell
 \ln\omega_\beta(e^{zN_F}),
 \label{eq:app-gibbs-mgf}
\end{equation}
where $\ln$ denotes the natural logarithm. Rigorous quantum large-deviation results establish that the scaled cumulant generating function $\phi_\beta(z)$ exists and is differentiable at the origin with $\phi'_\beta(0)=\rho_T^\beta>0$~\cite{Araki1969,Ogata2010}. Choosing $u>0$ sufficiently small ensures $\phi_\beta(-u)<0$, which yields the exponential bound $w_\ell^\beta\le C_\beta e^{-\gamma\ell}$ for constants $C_\beta,\gamma>0$. Expanding the information deficit for small $w$ gives
\begin{equation}
 1-f(1-w)=\frac w2\log_2\frac ew+O(w^2),
 \label{eq:app-reliability-asymptote}
\end{equation}
showing that in the thermal benchmark, a block length scaling as $\ell=O(\log(1/\delta))$ suffices to achieve $I_\ell\ge 1-\delta$ bits~\cite{Chernoff2007,Zwolak2014}.

Transferring this exponential reliability bound to unitary dynamics requires additional control over local equilibration. If a positive-density family of disjoint $\ell$-dimer blocks satisfies the trace-distance bound $\tfrac12\|\rho_{1,F}(t_L)-\rho_{\beta,F}\|_1\le\eta_{\ell,L}$, then
\begin{equation}
 w_\ell(F,t_L)\le C_\beta e^{-\gamma\ell}+\eta_{\ell,L}.
 \label{eq:app-thermal-transfer}
\end{equation}
If $\eta_{\ell,L}\to0$ uniformly as $L\to\infty$ for each fixed $\ell$, choosing a sufficiently large constant $\ell$ yields an extensive family of nearly complete records. This trace-distance equilibration condition can be stronger than our entropic criterion in Sec.~\ref{sec:theory}, which guarantees partial records without assuming local Gibbs thermalization.

For the canonical benchmarks in Table~\ref{tab:blocks}, $\beta_L$ is tuned at each system size to match the conserved seed energy $1.2J$, yielding $\beta_LJ\simeq0.180,0.216,0.247$ for $L=16,18,20$ via stochastic Lanczos imaginary-time evolution over eight random complex vectors. While both dynamical and thermal all-singlet probabilities drop steeply with $\ell$, a finite offset remains visible. Evaluating the full trace distance between dynamical reduced states at $L=20$, $Jt=60$ and their energy-matched Gibbs marginals gives approximately $0.04$ for two-dimer blocks and $0.15$ for four-dimer blocks. These diagnostics confirm substantial local relaxation while distinguishing finite-size behavior from idealized asymptotic thermalization.

\section{PXP clocks, local states, and readout robustness}
\label{app:pxp}
This appendix details the numerical protocol and basis-optimization checks supporting Fig.~\ref{fig:pxp}. We evolve the periodic Hamiltonian in Eq.~\eqref{eq:pxp} at $\lambda=0$ and $0.053$, where $Z_j=+1$ denotes an excited Rydberg state, $P_j=(1-Z_j)/2$ projects onto the ground state, and the reference N\'eel state $\ket{Z_2}$ has excited even sites. The seeded branch $X_{j_0}\ket{Z_2}$ de-excites site $j_0=2\lfloor L/4\rfloor$. Both branches evolve within the exact Rydberg-blockaded subspace (dimension $103682$ at $L=24$) without forward-scattering or tensor-network approximations.

\subsection{Reference-selected events and finite-size scope}
Stroboscopic observation clocks are determined exclusively from the unperturbed reference trajectory by locating peaks of $|m_{s,0}(t)|$ with a minimum temporal separation of $1/J$ and a minimum prominence of $0.005$, followed by refinement within windows of half-width $0.15/J$ on a fine grid of step $0.0025/J$. At each revival event, all sites and both branches are evaluated at the exact same time. Because ordinary and stabilized PXP use their own respective reference peaks, the revival index $n$ corresponds to slightly different physical times $\tau=Jt$ in the two models (Table~\ref{tab:pxp-clocks}).

\begin{table}[!htbp]
\caption{\label{tab:pxp-clocks}Summary of reference-selected revival events used in the PXP comparison, with dimensionless time $\tau=Jt$. These discrete stroboscopic events represent sampled revival peaks rather than continuous storage intervals.}
\centering
\begin{ruledtabular}
\begin{tabular}{rrrrr}
 & \multicolumn{2}{c}{Ordinary PXP} & \multicolumn{2}{c}{Stabilized PXP}\\
$L$ & events & final $\tau$ & events & final $\tau$\\
12 & 12 & 28.3900 & 11 & 27.2650\\
16 & 16 & 37.8175 & 15 & 37.1775\\
20 & 20 & 47.2650 & 19 & 47.0925\\
24 & 23 & 54.3475 & 22 & 54.5275
\end{tabular}
\end{ruledtabular}
\end{table}

Across all four chain lengths, our dataset comprises $138$ revival events. At $L=24$ and $\eps=0.25$ bits, stabilized PXP first reaches six fixed-$Z$ readers at $n=15$ ($\tau=37.1800$) and twelve readers at $n=20$ ($\tau=49.5725$), maintaining both counts at the subsequent revival. By contrast, ordinary PXP at $L=24$ never reaches either target within the simulated window and drops to zero quarter-bit readers from $n=7$ onward (although smaller rings such as $L=12$ exhibit finite-size revivals).

\subsection{Measurement optimization and its meaning}
Parameterizing the single-site marginals as $\rho_{b,j}=(1+\mathbf r_{b,j}\cdot\boldsymbol\sigma)/2$, we note that each application of the real Hamiltonian $H_\lambda$ flips the total excitation parity. Because both initial states have definite parity in the computational basis, time-evolved amplitudes in opposite parity sectors differ by a purely imaginary phase, which forces $\langle X_j\rangle_b=0$ identically at all times. Consequently, single-site Bloch vectors lie entirely within the $YZ$ plane, and projective readouts can be parameterized by the unit vector $\mathbf n(\theta)=(0,\sin\theta,\cos\theta)$ with $\theta\in[0,\pi)$ (since any out-of-plane component merely adds label-independent noise).

With conditional outcome probabilities $p_{b,j}(\theta)=[1+\mathbf r_{b,j}\cdot\mathbf n(\theta)]/2$, the single-site projective information along axis $\theta$ is
\begin{equation}
 I_j(\theta)=h_2\!\left(\frac{p_{0,j}+p_{1,j}}2\right)
 -\frac{h_2(p_{0,j})+h_2(p_{1,j})}{2},
 \label{eq:app-pxp-axis}
\end{equation}
where $\theta=0$ recovers the fixed-$Z$ readout $I_j^Z$ and maximizing over $\theta$ gives the site-optimized projective information $I_j^{\rm proj,*}=\max_\theta I_j(\theta)$. The measurement-independent single-site Holevo bound is given directly by the Bloch vector norms as
\begin{equation}
 \chi_j=h_2\!\left(\frac{1+|\bar{\mathbf r}_j|}{2}\right)
 -\frac12\sum_{b=0,1}h_2\!\left(\frac{1+|\mathbf r_{b,j}|}{2}\right),
 \label{eq:app-pxp-holevo}
\end{equation}
where $\bar{\mathbf r}_j=(\mathbf r_{0,j}+\mathbf r_{1,j})/2$, obeying $I_j^Z\le I_j^{\rm proj,*}\le\chi_j$.

To test sensitivity to the measurement basis at each revival event, we compare four reader counts, namely the fixed-$Z$ count $R^Z_\eps$, the uniform-axis count $R^{\rm unif}_\eps$ (which maximizes the number of readable sites over a single common angle $\theta$ shared by all observers at that event), the site-optimized projective count $R^{\rm proj}_\eps$, and the Holevo upper-bound count $R^\chi_\eps$. These obey $R^Z_\eps\le R^{\rm unif}_\eps\le R^{\rm proj}_\eps\le R^\chi_\eps$.

At the $\eps=0.25$-bit threshold, $R^{\rm unif}_{0.25}$ and $R^{\rm proj}_{0.25}$ coincide identically across all $138$ revival events, and at $L=24$ basis optimization adds at most two transient readers relative to fixed-$Z$ measurements without altering the late-time contrast between stabilized and ordinary PXP. Note that at the stabilized $n=12$ event ($L=24$), we find $(R^Z,R^{\rm unif},R^{\rm proj},R^\chi)=(0,0,0,5)$, where one site has $(I^Z,I^{\rm proj,*},\chi)=(0.22718,0.24591,0.29157)$ bits, showing that the Holevo bound can cross a threshold before any projective measurement does.

At the final stabilized $L=24$ revival ($n=22$), all four counts equal $24$ at $\eps=0.10$ bits, $12$ at $\eps=0.25$ bits, and $0$ at $\eps=0.50$ bits. At the final ordinary $L=24$ revival ($n=23$), all four counts vanish at all three thresholds, as the maximum single-site Holevo information has decayed to $0.09108$ bits (with a spatial mean of $0.05767$ bits). Thus, ordinary PXP loses its quarter-bit and tenth-bit records at late times even though the conditional marginals do not become strictly identical.

Finally, writing $\mathbf r_b=u_b\hat{\mathbf r}_b$ with relative angle $\vartheta$ decomposes the squared Bloch distance into radial and angular parts,
\begin{equation}
 |\mathbf r_0-\mathbf r_1|^2
 =(u_0-u_1)^2+2u_0u_1(1-\cos\vartheta).
 \label{eq:app-pxp-contrast}
\end{equation}
At the final stabilized $L=24$ revival, the reference Bloch lengths $u_0$ exceed $0.98$ on average across both sublattices, whereas the perturbed Bloch lengths $u_1$ average approximately $0.32$ and $0.59$ on the two sublattices, confirming that both differential mixedness and angular rotation contribute to the local contrast.

\subsection{Numerical checks}
Sitewise angular optimization is performed using grids of $361$ and $1441$ points followed by local bounded refinement, while uniform-axis counts are determined exactamente by intersecting the threshold-crossing angular intervals across all sites. Both angular resolutions yield identical integer counts across all $414$ combinations of system size, model, revival event, and threshold, with optimized single-site information agreeing to within $1.8\times10^{-15}$ bits and remaining invariant under $\pm10^{-9}$-bit threshold shifts. Independent cross-checks ofconstrained-basis partial traces against full Hilbert-space reductions at $L=12,16$ agree to within $1.1\times10^{-14}$ across all Bloch components.

\par
\end{document}